\documentclass[a4paper, 10pt, twocolumn, superscriptaddress]{revtex4-2}     
\usepackage{amsmath}
\usepackage{times}
\usepackage{graphicx}
\usepackage{hyperref}
\usepackage{setspace}
\usepackage{txfonts}
\usepackage{pxfonts}
\usepackage{float}
\usepackage{xcolor}
\usepackage{comment}
\usepackage{algpseudocode}
\usepackage{adjustbox}
\usepackage[toc,page]{appendix}
\usepackage{array}
\newcolumntype{x}[1]{>{\centering\arraybackslash\hspace{0pt}}p{#1}}
\newcolumntype{R}[1]{>{\raggedleft\let\newline\\\arraybackslash\hspace{0pt}}m{#1}}
\newcolumntype{L}[1]{>{\raggedright\let\newline\\\arraybackslash\hspace{0pt}}m{#1}}
\usepackage{ulem}
\usepackage{natbib}
\usepackage{nameref}
\usepackage{soul}

\let\emph\textit

\usepackage{hyperref}
\begin{document}


\title{Fast degree-preserving rewiring of complex networks}
\author{Shane Mannion}
	\affiliation{Department of Applied Mathematics and Computer Science, Technical University of Denmark, 2800 Kongens Lyngby, Denmark.}
\author{P\'adraig MacCarron}
	\affiliation{MACSI, Department of Mathematics and Statistics, University of Limerick, Limerick, V94 T9PX, Ireland}
\author{Akrati Saxena}
        \affiliation{Leiden Institute of Advanced Computer Science, Leiden University, Niels Bohrweg 1, 2333CA, Leiden, The Netherlands}
\author{Frank W. Takes}
        \affiliation{Leiden Institute of Advanced Computer Science, Leiden University, Niels Bohrweg 1, 2333CA, Leiden, The Netherlands}

\begin{abstract}
{In this paper we introduce a new, fast, degree-preserving rewiring algorithm for altering the assortativity of complex networks, which we call the \textit{Fast total link (FTL) rewiring} algorithm. Commonly used existing algorithms require a large number of iterations, particularly in the case of large dense networks. This can especially be problematic when we wish to study ensembles of networks. In this work we aim to overcome aforementioned scalability problems by developing an algorithm which improves upon existing algorithms. The method we propose performs better than existing methods by several orders of magnitude for a range of structurally diverse complex networks, both in terms of the number of iterations required, and time taken to reach a given assortativity value. Additionally, we highlight the result that we can rewire a graph to have the maximum (minimum) possible assortativity for the given degree sequence while having just one component.
}
\end{abstract}

\maketitle

\section{Introduction}
\label{section:intro}
Rewiring of networks is an important task that is of broad interest to the network science community. We rewire a network to change one property of the network either with no regard for other properties (random rewiring) or while holding other properties fixed, for example the degree sequence. We refer to the latter as degree-preserving rewiring. Often, we use rewiring algorithms to create networks with desirable properties for the study of dynamical processes on the networks. Examples of such rewiring algorithms include rewiring to increase network robustness~\cite{chan2016rewiringforrobustness}, clustering~\cite{alstott2019localclustering, kashyap2018mechanisms}, synchronisability~\cite{bertotti2020rkplane} and assortativity~\cite{kashyap2018mechanisms, van2010influence, winterbach2012greedy}. In this paper we focus on degree-preserving rewiring for tuning assortativity values. With such algorithms, we take a graph and a target assortativity value as inputs, and return a graph where the node degrees are unchanged, but the assortativity value is now at or very close to the target value. 

Broadly speaking, rewiring algorithms are either Monte-Carlo based, where we randomly select edges to rewire repeatedly until we achieve our goal, or simulated annealing based~\cite{bertsimas1993simulatedannealing}. In simulated annealing, we define a ``system temperature''. Typically we select a large value for this temperature and we decrease it by some chosen factor at each iteration of the algorithm. The higher the temperature, the larger the probability that we allow a backward step (i.e., a step that moves away from the goal value)~\cite{de2018fundamentals}. Allowing these backwards steps helps us to avoid local optima. Both approaches can be used to the same end; for example in ~\cite{chan2016rewiringforrobustness} and ~\cite{buesser2011optimizingannealing}, the aim is to increase the robustness of networks, with the former employing Monte-Carlo based algorithms, and the latter using simulated annealing. In this paper, we focus on Monte-Carlo based degree-preserving rewiring algorithms.
Both Monte-Carlo and simulated-annealing based algorithms are typically not designed with computational efficiency in mind. The majority of algorithms that exist in the literature work by rewiring just two edges at a time~\cite{kashyap2018mechanisms, van2010influence, winterbach2012greedy}. The contribution of one edge to the overall assortativity is generally small. It is no surprise then that large numbers of iterations of algorithms that rewire just two edges at once are required to achieve a given rewiring goal. 

The approach we propose in this work is to rewire a larger number of edges at once, reaching the target assortativity values much faster. In a simulated annealing algorithm, we create edges randomly. These new edges may either increase or decrease the assortativity. If the new edges decrease the distance to our target assortativity, we add them with probability $1$. If they increase the distance to our goal, we add them with probability $p$. This ``backwards step'' helps to avoid local maxima and minima. In our algorithm, we create edges strategically instead of randomly. It is for this reason that we use a Monte-Carlo based approach. A backwards step, as utilised in simulated annealing is (in the vast majority of cases) not possible, due to the fact that we construct new edges in such a way that they will increase the assortativity.  We demonstrate the efficacy of the approach by studying both random graphs with varying degree distributions and a number of real-world networks.

The remainder of this paper is laid out as follows: In Section~\ref{section:background} we discuss the widely used existing algorithm and its limitations. In Section~\ref{section:improvement} we discuss how our algorithm works for both assortative and disassortative rewiring. Following discussion of our algorithm, we compare its performance to that of the original algorithm using simulated networks and real-world networks in Section~\ref{section:results}.Finally, we discuss conclusions and potential future work in Section~\ref{section:conclusions}.
\section{Background \& Original rewiring algorithm}
\label{section:background}
A graph or network is an object comprised of a set of $N$ nodes and a set of $L$ links (or edges) that connect these nodes. The number of edges that are connected to a node is called the \textit{degree} of a node and is denoted $k$. The \textit{degree distribution} $p_k$ of a graph (i.e., the distribution of the probability that a randomly chosen node has degree $k$), is the subject of much research~\cite{albert2002statistical, amaral2000classes, newman2001random}. In addition, the \textit{assortativity} of a graph, which is the correlation between the degrees of nodes that are connected by edges, has important practical consequence, such as impacting the rate at which a disease will spread across a network~\cite{newman2002assortative}. We can vary the latter of these properties while keeping the former fixed through the use of a degree-preserving rewiring algorithm.

\begin{figure}[t]
    \centering
    \includegraphics[page=3, width=0.45\textwidth]{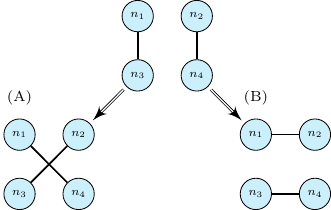}
    \caption{The two ways in which a pair of edges can be rewired via the algorithm described in Section~\ref{section:background}}.
    \label{fig:originalAlgorithm}
\end{figure}

First introduced in~\cite{maslov2002originalrewiring}, the degree-preserving rewiring algorithm works as follows. We select two edges, ($a, b$) and ($c, d$). We consider the ways in which to configure two edges among these four nodes, i.e., ($a, c$) and ($b, d$) or ($a, d$) and ($b, c$). If our goal is to increase the assortativity, then we select a pair of edges that results in an increase in the assortativity. If we want to induce disassortativity, then we select a pair of edges that reduces the assortativity. In Figure~\ref{fig:originalAlgorithm}, we demonstrate the two ways in which a selected pair of edges can be rewired. The choice of situation (a) or (b) (or the original) is based on the formula for the degree assortativity, denoted by $r$ described in~\cite{van2010influence},
\begin{equation}
    r = 1 - \frac{\sum_{e} (k_i - k_j)^2}{\sum_{i = 1}^{N} k_i^3 - \frac{1}{2L}\left(\sum_{i = 1}^N k_i^2\right)^2}.
    \label{eqn:vanmeighemassort}
\end{equation}
In this equation, $k_i$ is the degree of node $i$, and the subscript $e$ indicates we are summing over all edges. The authors~\cite{van2010influence} note that a degree-preserving rewiring will only change the numerator of the fraction in Equation~(\ref{eqn:vanmeighemassort}). When we rewire two edges, there are three possible pairs of edges which we can form (including the original pair). We show these in Figure~\ref{fig:originalAlgorithm}. A similar figure appears in~\cite{SHANG201749}. A successful rewiring requires that the edges we wish to form do not already exist in the graph. In the scenario in which $k_{n_1} \ge k_{n_2} \ge k_{n_3} \ge k_{n_4}$ - where $k_{n_i}$ is the degree of node $i$ - in Figure~\ref{fig:originalAlgorithm}, we would have $r_{(\rm B)} \ge r \ge r_{( \rm A)}$, where $r$ is the assortativity of the graph before the rewiring step and $r_{(\rm A)}$ and $r_{(\rm B)}$ are the assortativity values after rewiring to configuration $(\rm A)$ and $(\rm B)$ respectively. If we want to increase the assortativity, we would choose configuration $(\rm B)$, and if we wish to decrease the assortativity we would choose configuration $(\rm A)$. 

Much research on rewiring algorithms pertains to the development of algorithms that produce maximally assortative graphs, or ``good'' results~\cite{stokes2018common, winterbach2012greedy, meghanathan2016maximal}. As noted by the authors of~\cite{stokes2018common}, common algorithms often fail to provide the graph with maximum assortativity for a given degree sequence, which they prove by finding counter examples in each case. 

In the their search for an algorithm to produce a maximally assortative graph, the authors of~\cite{winterbach2012greedy} state that is unknown how close the assortativity of a graph produced by a rewiring algorithm is to the true maximum or minimum assortativity for a given graph. This brings us to one of the primary results of this paper. Here, we highlight the Havel-Hakimi algorithm~\cite{hakimi1962realizability, kashyap2018mechanisms}. This result, which seems to have gone unnoticed in rewiring research until now, provides an answer to many of the research questions posed by work in this field, and forms the basis of the rewiring algorithm we propose.

The Havel-Hakimi algorithm allows one to test if a given degree sequence is graphical by forming edges in specific manner, which we describe in Section~\ref{sec:ftlalg}. If the edges can be added in the manner they describe, then the sequence is graphical. What's more, the graph created by performing this algorithm is both connected and has the maximum possible assortativity for that degree sequence.
\section{Proposed Method: Fast Rewiring Algorithm}
\label{section:improvement}
In this section we first discuss an initial, modest improvement we made to the existing algorithm. The preliminary results obtained ultimately lead to the workings proposed algorithm outlined in Section~\ref{sec:ftlalg}.
\subsection{Naive simultaneous rewiring}
\label{sec:naiverewiring}
We begin by introducing an additional parameter to the existing algorithm, $m$. We define the additional parameter $m$ as the number of edges to be rewired in a given iteration of the algorithm. To the authors' knowledge, we are the first to propose selecting more than two edges at once to rewire. Existing algorithms take a graph $G$, and a target assortativity $r_t$ as inputs. 

The logic of this rewiring strategy is the same as that of the original algorithm discussed in Section~\ref{section:background}; however, the formulation is based on a different derivation. As we have shown in previous work~\cite{maccarron2023correlation}, the assortativity can be written as
\begin{equation}
    r = \frac{1}{2L} \sum_{i,j}  \frac{A_{ij}(k_i - \text{E}[\kappa]) (k_j -\text{E}[\kappa] ) }{\text{E} [\kappa^2] - \text{E}[\kappa]^2},
    \label{eqn:maccarronassortativity}
\end{equation}
where $A$ is the adjacency matrix and E$[\kappa]$ is the average degree at the end of an edge (not to be confused with the mean degree of the network).~It follows from Equation~(\ref{eqn:maccarronassortativity}) that if two nodes are such that one has a degree greater than the average degree at the end of an edge and the other has a degree less than the average degree at the end of an edge, this edge will contribute negatively to the assortativity. If both nodes have degree greater than or both have degree less than the average degree at the end of an edge, this edge will contribute positively to the assortativity. The change we make exploits this fact.

The algorithm works as follows: we choose $m$ edges, which connect $2m$ nodes which \textit{need not} be unique. We rank the nodes by their degree from $0$ to $2m-1$. If we wish to increase the assortativity, then we connect node $0$ to node $1$, node $2$ to node $3$, and so on, until node $2m - 2$ is connected to node $2m - 1$. This way, we remove one edge and add one edge to each node. We add the new edges to the graph, and remove the $m$ edges that we chose at the beginning of the iteration. If we wish to decrease the assortativity, we form the new edges between nodes $0$ and $2m-1$, nodes $1$ and $2m-2$, and so on.

\begin{figure}[h]
    \centering
    \includegraphics[page=5, width=0.5\textwidth]{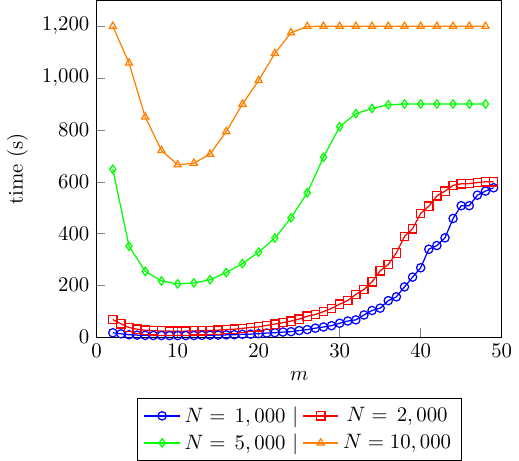}
    \caption{Time taken to rewire Erd\H{os}-R\'enyi graphs of varying sizes to an assortativity value of 0.25 for different numbers of chosen edges $m$ per algorithm iteration.}
    \label{fig:originalimprovement}
\end{figure}

Our preliminary experiments showed that a rewiring algorithm based on the procedure described above showed modest success, although it was inconsistent.~In some cases, it reduced the amount of time taken to rewire a graph by up to 95\%, in other cases it made no improvement whatsoever. In Figure~\ref{fig:originalimprovement} we show the time taken to rewire Erd\H{o}s-R\'enyi graphs of varying sizes, all with an average degree of 5, from their original assortativity values ($r \approx 0$) to an assortativity value of 0.25. As we can see, even though increasing the number of edges rewired at each step reduces the time taken (before it increases again, which will be discussed later in this section), the amount of time taken to rewire the graphs is still quite large. 

The lackluster performance of this method comes from the fact that any of the edges we attempt to add may already exist in the network. If one (or more) of the edges we wish to add does exist, the sample of edges must be discarded and redrawn. Similarly, if the set of potential edges contains a self edge it must also be discarded.~As a result, increasing the number of edges rewired in a given iteration does not necessarily improve the algorithm. To illustrate why, we look at the adjacency matrix.

Consider a network with $N$ nodes and $L$ edges. Allowing no self edges, the adjacency matrix of this network has $2L$ non-zero elements. If we remove $m$ edges from the graph in our rewiring problem, then the proportion of non-zero elements in the adjacency matrix is 
\begin{equation}
    \frac{2L - 2m}{(N-1)^2}.
\end{equation}
We then must add $m$ edges back into the graph, i.e., add $2m$ ones to the adjacency matrix. Since we are preserving the degree, these ones must be added to the same portion of the adjacency matrix that they were removed from, i.e., the square submatrix made up of the rows and columns corresponding to the (up to) $2m$ nodes affected by the removal of our $m$ edges. We then make the following two assumptions. First, we assume that the ones in the adjacency matrix are randomly distributed (which is only the case for an Erd\H{o}s-R\'enyi random graph, however, this simple case is sufficient to illustrate our point). As such, the proportion of ones in this submatrix is the same as that for the adjacency matrix.~Secondly, we assume that we are adding edges randomly and not based on the degrees of the nodes. Essentially, we must randomly select $m$ pairs of elements of this submatrix (one for each node of the edge being formed). If any selected element is equal to one, we must discard the sample.

This process can be viewed a Bernoulli trial, and our probability of success for a sample size of $m$ edges is 
\begin{equation}
    p = \left(1 - \frac{2L - 2m}{(N-1)^2}\right)^m. \label{eqn:probsuccess}
\end{equation}
In testing, we find that a small increase in $m$ improves the algorithm, as the probability of success remains relatively high.~In fact, in the limit of large $N$, $p$ is essentially $1$ when $m=2$~\cite{xulvi2004reshuffling}. As such, a small increase in $m$ does not significantly reduce the probability of success.~Once $m$ becomes larger, the probability of success becomes very small.~We observe this in Figure~\ref{fig:originalimprovement}, where we can see that increasing the number of edges selected per iteration helps only to a certain extent.

\subsection{Fast total link rewiring}
\label{sec:ftlalg} 
\begin{figure}[t]
    \centering
    \includegraphics[page=10, width=0.45\textwidth]{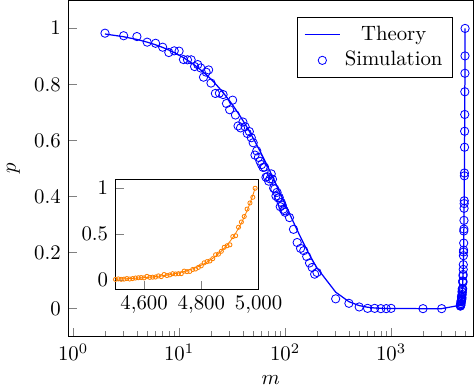}
    \caption{Probability that a sample of $m$ rewired edges will be accepted vs $m$, for an Erd\H{os}-R\'enyi graph of size $N=1,000, L=4,928$. Inset shows $p$ versus large values of $m$ on a linear scale. Note that $p$ and $m$ here are not to be confused with the parameters of the $G(n, M)$ and $G(n, p)$ models.}
    \label{fig:probsuccess}
\end{figure}

It is the inspection of Equation~(\ref{eqn:probsuccess}) that leads us to the solution of the problem discussed in the previous section.~In this equation, the probability of success depends on the number of edges being sampled at once, but it does so in a non-linear way. In Figure~\ref{fig:probsuccess} we plot the probability of success $p$ versus the number of edges being sampled at once $m$. For small values of $m$, $p$ remains high. For intermediate values of $m$, $p$ is very small. As $m$ approaches $L$, the probability of success increases again. Finally, in the case where $m = L$, the probability of an edge that we are attempting to form existing in the graph is zero, as we have first removed all edges from the graph. With this in mind, we can describe fully our new degree-preserving rewiring algorithm.
\subsubsection{Step one - Havel-Hakimi algorithm}
\label{fastassortativerewiring}
Our new algorithm, which we call \textit{fast total link rewiring} (or \textit{FTL} rewiring), has two main steps. First, we increase (decrease) the assortativity to a very high (low) value by rewiring all edges at once. We refer to this step as \textit{total rewiring}.~From this extreme value, we rewire $m$ edges at once as described in Section~\ref{sec:naiverewiring} to reach the target value.

\begin{figure}[t]
    \centering
    \includegraphics[page=53, scale = 1.2]{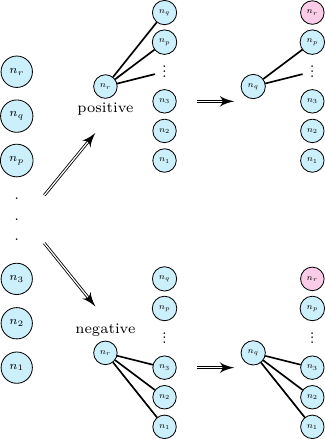}
    \caption{The total rewiring step of the FTL rewiring algorithm described in section~\ref{section:improvement}. Magenta coloured nodes represent those nodes that have been connected to a number of other nodes equal to their degree in the original graph.}
    \label{fig:OurStepOne}
\end{figure}

To increase the assortativity of a graph, we begin with a total rewiring. In this total rewiring step, we remove all edges, and the nodes in the network are ranked by their original degree.~We connect the highest degree node (with say degree $k$) to the next $k$ highest ranking nodes in the graph by degree.~We perform the same procedure for the second-highest degree node in the graph, and continue like so, ensuring that we are at no point exceeding the degree of any node. In Figure~\ref{fig:OurStepOne}, we show a diagram of this process. This step of the algorithm is in fact the Havel-Hakimi algorithm~\cite{hakimi1962realizability, havel1955remark}, which has been proven to create the graph of highest possible assortativity. In the disassortative case, we reverse this process. After ranking the nodes by degree, we connect the lowest degree node (with degree $k$), with the $k$ highest ranking nodes in the graph.
\subsubsection{Step two - rewiring $m$ edges at once}
\begin{figure}[t]
    \begin{center}      
    \includegraphics[page=1, width=0.45\textwidth]{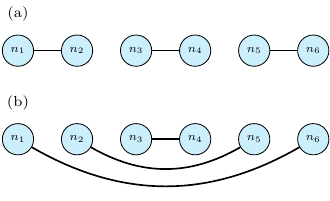}
    \end{center}
    \caption{(a): The way in which nodes are rewired using the FTL algorithm to increase assortativity. (b): The way in which nodes are rewired using the FTL algorithm to decrease assortativity. In both cases nodes are ranked such that $k_{n_i} \le k_{n_i + 1}$. All algorithms described in section \ref{section:improvement}.}
    \label{fig:ourAlgorithm}
\end{figure}

Following our total rewiring, we want to decrease (increase) the assortativity from this maximum (minimum) value to whatever the desired value is. To do this we select two (or $m$) edges from our rewired graph, and again order the nodes (which now need not be unique) by their degree. We connect the node with the largest degree to the node with the smallest degree, and connect the middle two nodes.~We repeat this process until the desired assortativity value is reached. This is the same process that is described in our discussion of the simultaneous rewiring in Section~\ref{sec:naiverewiring}. Note also that we check these new edges to ensure we are not attempting to add any loops, duplicate edges, or edges that are already present in the graph. If any edge fails this check, then we discard the sample of $m$ edges and draw a new sample. In Figure~\ref{fig:ourAlgorithm}~(a) we show how edges are rewired in this step of the algorithm.

Earlier in this section, we discussed how increasing the number of edges being rewired at once does not necessarily improve the speed of the rewiring. This is no longer the case.~As we have already mentioned, if an edge connects two nodes whose degrees are both greater (or less) than than the average degree at the end of an edge for the graph, such an edge will increase the assortativity. After applying the Havel-Hakimi algorithm, the vast majority of the edges in the graph will contribute positively to the assortativity, and since we are now trying to form disassortative edges, we can rewire many more edges at each iteration without running into the problem of these edges already being present in the graph. This allows us to rewire a larger sample of edges at once, further improving the speed of the FTL rewiring algorithm. 
\label{fastdisassortativerewiring}
\begin{figure*}[t]
    \centering
    \includegraphics[page=11, width=0.3\textwidth]{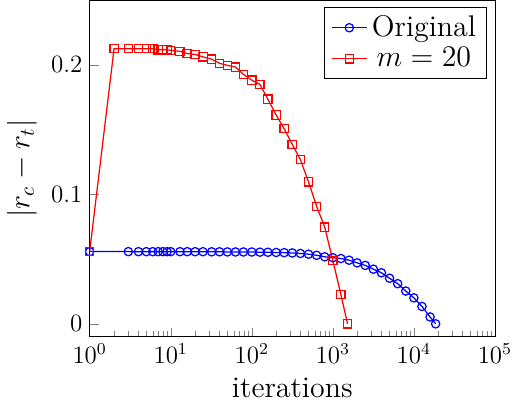}
    \includegraphics[page=12, width=0.3\textwidth]{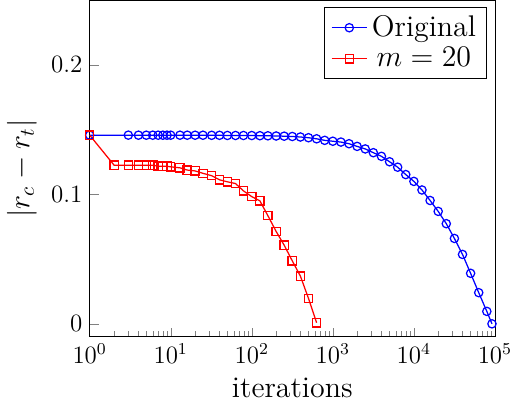}
    \includegraphics[page=13, width=0.3\textwidth]{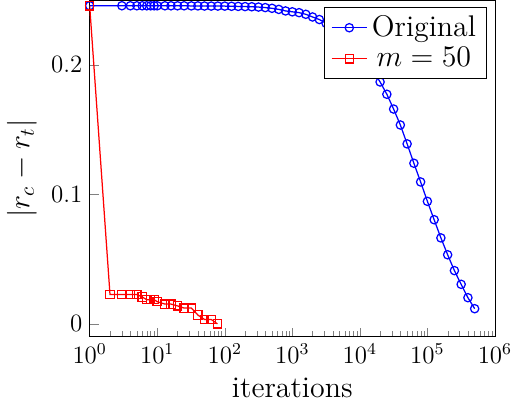}
    \includegraphics[page=14, width=0.3\textwidth]{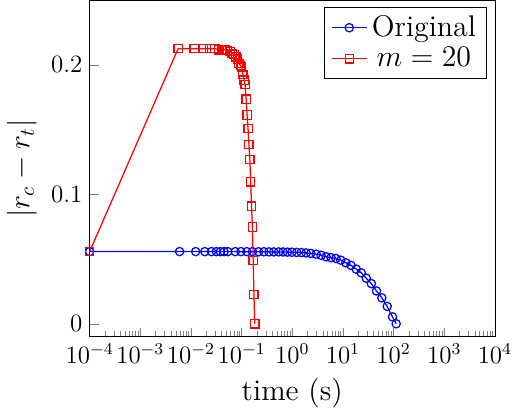}
    \includegraphics[page=15, width=0.3\textwidth]{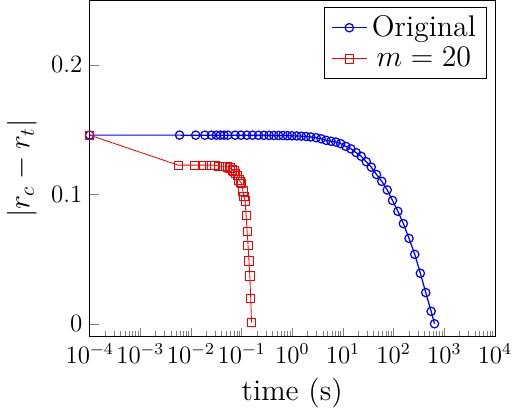}
    \includegraphics[page=16, width=0.3\textwidth]{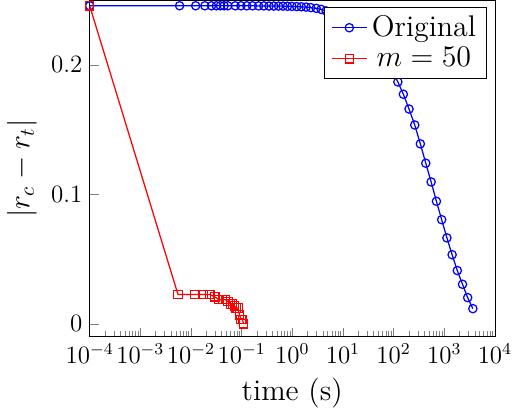}
    \caption{Difference between the current assortativity at a given iteration (top) or time (bottom) and the target assortativity for the US airport network. The target assortativity values are 0.01 (left), 0.1 (middle), and 0.2 (right). Blue lines represent the original algorithm, and red points represent the FTL algorithm with the best performing choice of $m$. In the leftmost plots we see a drastic increase in $|r_c - r_t|$ before it then reduces to zero. This is due to the fact that our initial assortativity is $\approx-0.05$, the assortativity after the first iteration of the FTL algorithm is $\approx 0.22$, and our target is 0.01, resulting in the large increase in $|r_c - r_t|$.}
    \label{fig:airportresults}
\end{figure*}

Pseudocode corresponding to our algorithm can be found in the~\hyperref[section:appendix]{Appendix}. Having described our FTL rewiring algorithm, we next move on to showing how it performs when compared to the original rewiring algorithm.
\section{Results}
\label{section:results}
In this section we 
look at the results of our testing of the FTL rewiring algorithm.~In particular we wish to look at four aspects of the algorithms performance.~Most importantly, after outlining the experimental setup in Section~\ref{sec:expsetup}, we compare the performance of the FTL rewiring to the original rewiring algorithm. This is covered in Section~\ref{results:speed}. Additionally, in Section~\ref{section:mandr} we look at how sensitive the FTL rewiring algorithm is to the choice of the parameter $m$, and how the target assortativity, $r_t$, affects the performance. In Section~\ref{results:largegraphs} we study the performance of the FTL algorithm on a large graph.

\subsection{Experimental Setup}
\label{sec:expsetup}
To test the FTL rewiring algorithm, we employ it on both real-world network datasets, and randomly generated graphs. In both cases, we select several target assortativity values to rewire each network to. The reason we pick several is that due to the first step (i.e., the total rewiring step) of the FTL rewiring algorithm which we described in the previous section, rewiring the graph to assortativity value close to its maximum or minimum will be, somewhat counterintuitively, faster than rewiring it to a value close to its starting value. This is not the case for the original algorithm. As such, it may be the case that for small changes in assortativity value, the original algorithm performs better and we will explore this is the coming sections.



We test the FTL rewiring algorithm's performance on five empirical networks. Namely we look at the US airport network used by Van Meighem \textit{et al.}~in~\cite{van2010influence}, three subsets of the friendship network from the Deezer music streaming service taken from three European countries; Romania (RO), Hungary (HR) and Croatia (HU)~\cite{deezer}. Finally, we look at the Dogster social network~\cite{konect:all} in Section~\ref{results:largegraphs}. We chose the airport network as it has been previously studied in the context of rewiring algorithms. We chose the Deezer networks as they each have a similar number of nodes ($\approx 50,000$), but very different edge counts, resulting in different densities and average degrees, and we chose Dogster as an example of a large network ($\approx 250,000$) nodes. Properties of these networks are shown in Table~\ref{tab:datasets}.

In the case of the airport network, we show both the number of iterations required and the time required for the rewiring. In the case of the other networks mentioned above, we show only the time as it is the more important metric. In all cases we show only the results for our best-performing choice of edges to rewire at once $m$, with the exception of Figure~\ref{fig:airportsamplesizecomparison} where we study the variation in performance with choice of $m$.

In addition to the real world networks described above, we study the performance of the FTL rewiring algorithm on randomly generated networks with degree sequences drawn from the following distributions; Poisson (Erd\H{o}s-R\'enyi graphs), exponential, lognormal, Weibull, and Barab\'asi-Albert graphs. We generate the Poisson and Barab\'asi-Albert graphs using the Networkx Python package's built in methods~\cite{networkx}. We generate graphs from the other distributions using the configuration model~\cite{newman2018networks}, with degree sequences drawn from discretised versions of the distributions (see Reference~\cite{mannion2023robust} for more information on discretising these distributions) using inverse transform sampling. All of the graphs discussed in this section have $N=5,000$ nodes and an average degree $\langle k \rangle \approx 5$, to allow for a fair comparison between graphs of differing distributions.

\subsection{Speed of the FTL rewiring algorithm}
In this section we focus on the improvement in speed obtained by implementing the FTL algorithm. First we look at this effect on real-world networks and then on randomly generated networks.
\label{results:speed}
\subsubsection{Real-world networks}
\begin{table}[b]
\begin{tabular}{|l|r|r|r|r|r|}
\hline
\multicolumn{1}{|c|}{\textbf{Network}} & \multicolumn{1}{c|}{\textbf{$N$}} & \multicolumn{1}{c|}{\textbf{$L$}} & \multicolumn{1}{c|}{\textbf{$r$}} & \multicolumn{1}{c|}{$\langle k \rangle$} & \multicolumn{1}{c|}{source} \\ \hline
Airport                                & \,\,\,2,179                             & \,\,\,31,326                           & \,\,\,-0.045                            & 28.75 & \cite{van2010influence}                 \\ \hline
Deezer (RO)                           & 41,773                             & 125,826                           & 0.11  & 6.02                           & \cite{deezer}                                   \\ \hline
Deezer (HU)                               & 47,538                            & 222,887                           & 0.207   & 9.38                      & \cite{deezer}                                   \\ \hline
Deezer (HR)                               & 54,573                            & 498,202                            & 0.197    & 18.26                      & \cite{deezer}                                   \\ \hline
Dogster                                & 260,390                            & 2,148,179                          & -0.09   & 16.50                     & \cite{konect:all}                                   \\ \hline
\end{tabular}

\caption{Properties of the chosen real-world networks. $N$ is the number of nodes, $L$ the number of edges, $r$ is the initial assortativity for each graph, and $\langle k \rangle$ is the average degree of the graph.}
\label{tab:datasets} 
\end{table}
An overview of datasets considered is given in Table~
\ref{tab:datasets}. 
The first network we test the algorithm on is the airport network which has $N = 2,179$ nodes and $L = 31,326$ edges.~In this network nodes represent airports and edges represent flights between airports. In Figure~\ref{fig:airportresults} we show the performance of both the FTL rewiring algorithm and the original described by Van Meighem \textit{et al.}~in~\cite{van2010influence}. Performance here is measured in terms of the distance between the current assortativity of the graph ($r_c$) and the assortativity that we are trying to reach ($r_t$). As such, the faster each curve goes to zero, the better. The red curve in each panel is the best performing sample size for our algorithm for each target. 

Note that the initial assortativity of the graph is approximately $-0.05$. As such, moving from left to right across the Figure, the initial distance between $r_c$ and $r_t$ increases. As we can see, the FTL algorithm requires several orders of magnitude less time and fewer iterations to achieve the same result as the original. 

\begin{figure*}[t]
    \centering
    \includegraphics[page=77, width=0.3\textwidth]{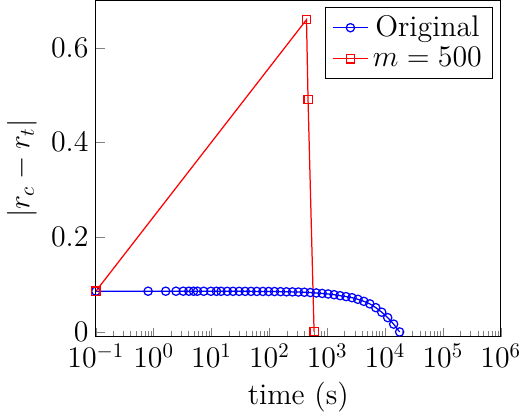}
    \includegraphics[page=89, width=0.3\textwidth]{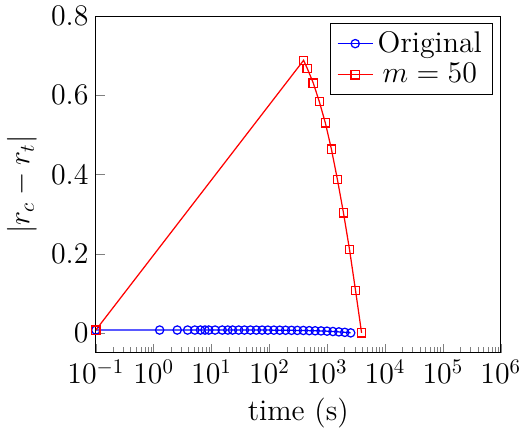}
    \includegraphics[page=81, width=0.3\textwidth]{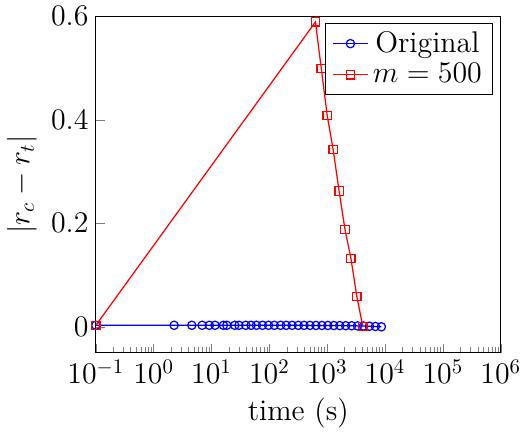}
    \includegraphics[page=78, width=0.3\textwidth]{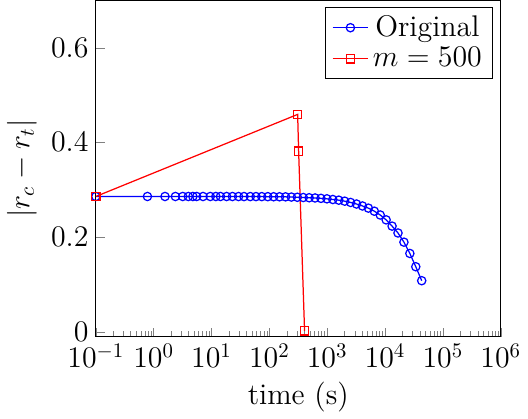}
    \includegraphics[page=90, width=0.3\textwidth]{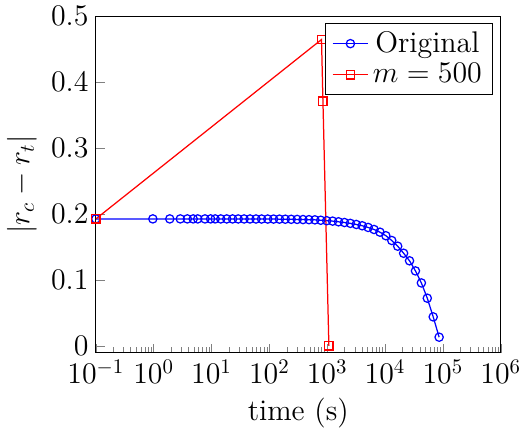}
    \includegraphics[page=82, width=0.3\textwidth]{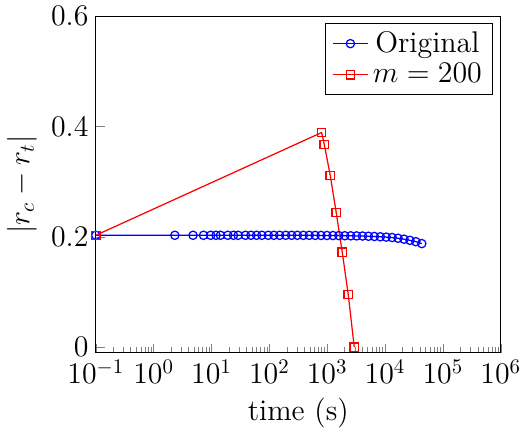}
    \caption{Difference between the current assortativity at a given time and the target assortativity for the different samples of the Deezer network; Romania (left), Hungary (middle), and Croatia (right). The target assortativity values are 0.2 (top) and 0.4 (bottom). Blue lines represent the original algorithm, and red points represent the FTL algorithm with the best performing choice of $m$. Note that we do not plot every time point to avoid visual clutter, hence the few markers for some curves. Note also in the top middle and top right panels the y-axis range is slightly altered for clarity. The original rewiring algorithm was stopped after 12 hours for the Romania and Croatia networks, and after 24 hours for the Hungary network. Hence, not all curves reach zero.}
    \label{fig:deezer_positive}
\end{figure*}

Next, we look at the Deezer network datasets. We note that the optimal choice of $m$ varies across the three datasets. We again highlight the dramatic improvement in rewiring time achieved by our algorithm. Looking at the top-middle and top-right panels of Figure~\ref{fig:deezer_positive}, the times for each algorithm appear comparable, with the original algorithm being faster for rewiring the subgraph corresponding to Hungary. We note here the original assortativity value for this graph is $r = 0.207$, and the target is $r = 0.2$. So in a comparable amount of time, the original algorithm spans the range of assortativity values from $0.2$ to $0.207$. Meanwhile, our algorithm spans the entire possible range of values form our target of $0.2$, to the maximum possible assortativity, $0.865$. Hence, for any other target value, our algorithm will outperform the original, as can be seen for the target of $0.4$ on the same graph. In all other cases in Figure~\ref{fig:deezer_positive}, our algorithm improves upon the original by several orders of magnitude.

In the next section, we discuss the maximum assortativity value we observe for different networks in more detail and so we use this opportunity to again highlight the dramatic improvement that our algorithm offers over the original in terms of both time taken to rewire a network and the required number of iterations.

Going back to the US airport network, we see a reduction in time taken to rewire the network from $\approx3 \times 10^3$ seconds to $\approx0.1$ seconds, a reduction of over 99.99\% in terms of time, and we see similar reductions for the Facebook network. In fact, somewhat as expected, we see that the larger the distance between the current assortativity and the target assortativity, the better our method performs. With the original algorithm only outperforming our method in the case where the desired change in assortativity is smallest, the difference in performance is also minimal. 
\subsubsection{Randomly-generated graphs}
One of the results we find in our study of randomly generated graphs is that the maximum and minimum assortativity values that one can obtain for a given network differ depending on the distribution from which the node degrees are drawn. As a consequence of this, we use the total rewiring step of the FTL rewiring algorithm to find maximum and minimum assortativity values for a handful of graphs from each distribution, and then use these to choose reasonable target values.
As we can see in Figure~\ref{fig:RandomGraphs}, the results are very strong. In all cases we reduce the time taken to rewire the graphs by several orders of magnitude, with the corresponding reduction in the number of iterations taken being even more dramatic. We see that in most cases, a choice of large $m$ is favorable, which is likely due to the relatively low densities of these networks. This relationship is discussed more in the next subsection.
\subsection{Impact of $m$ and $r_t$}
\label{section:mandr}
In the experiment shown in Figure~\ref{fig:airportresults}, when rewiring to a target value of 0.01 or 0.1, rewiring 20 edges at once is the best choice for this network. For a target value of 0.2, however, rewiring 50 edges at once is a better choice. The reasoning for this is straightforward. For the airport network, the total rewiring step of the FTL rewiring algorithm increases the assortativity to 0.22. In Section~\ref{section:improvement} we mentioned that once we have performed the total rewiring step, the probability of attempting to create an edge that already exists in the network is very small. However, as we perform more iterations, this probability increases as more disassortative edges exist. Furthermore, the more edges we rewire at once, the more likely we are to attempt to form one of the relatively few disassortative edges that exist in the network. In the case of the target values $0.1$ and $0.01$, the choice of a large value of $m$ results in the probability of failure being large enough that a large proportion of samples of edges fail. In the case of the target value $0.2$, however, $m=50$ is the superior choice. This is because the number of iterations required to go from the an assortativity value of $0.22$ to $0.2$ is sufficiently few that that the large change in assortativity per iteration outweighs the increased probability of failure.
\begin{figure*}[h]
    \centering
    \includegraphics[page=54, width=0.45\textwidth]{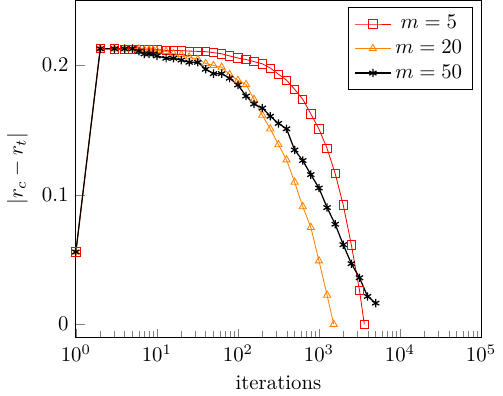}
    \caption{Performance of the FTL algorithm with sample sizes $m=5, 20,$ and $50$. The $m=50$ case starts off the fastest but as the number of iterations increases, it performs worse due to an increasing number of failed iterations.}
    \label{fig:airportsamplesizecomparison}
    \includegraphics[page=91, width=0.45\textwidth]{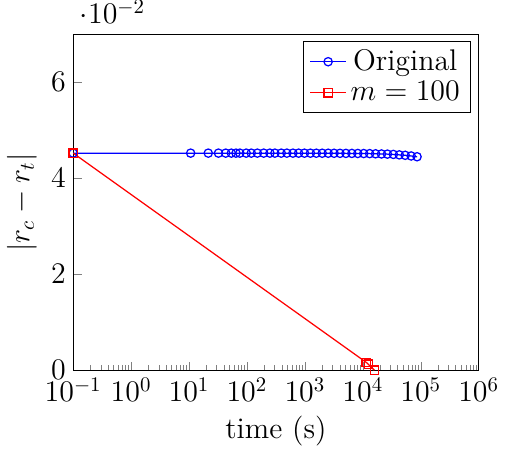}
    \caption{Difference between the current assortativity at a given iteration and the target assortativity for the Dogster network. Note the y-axis is scaled by 100. Rewiring target is -0.05 from a starting value of -0.09. Experiment for the original algorithm was cut off after 24 hours.}
    \label{fig:dogster_positive}
    \centering
    \includegraphics[page=92, width=0.45\textwidth]{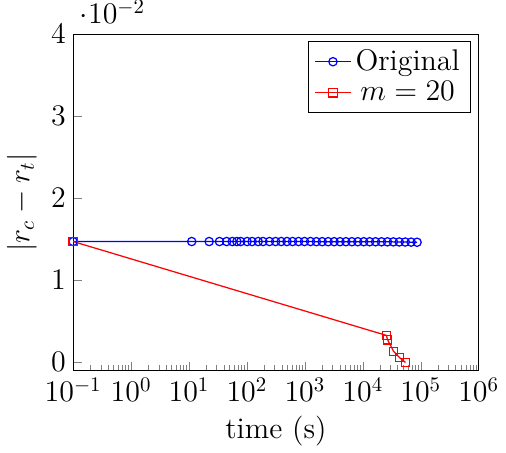}
    \caption{Difference between the current assortativity at a given iteration and the target assortativity for the Dogster network. Note the y-axis is scaled by 100. Rewiring target is -0.11 from a starting value of -0.09. Experiment for the original algorithm was cut off after 24 hours.}
    \label{fig:dogster_negative}

\end{figure*}

To illustrate the effects of different $m$ values described above, in Figure~\ref{fig:airportsamplesizecomparison}, we show the performance of different sample sizes in rewiring the airport network. As we can see, initially the $m = 50$ curve is decreasing the fastest, but as the number of iterations increases, its performance deteriorates as described above.

The optimal choice of $m$ is related to the edge density of the graph. If we again look at Equation~(\ref{eqn:probsuccess}), the probability of success depends on $\frac{2M - 2m}{(N - 1)^2}$. Evaluating the full equation for the power-grid network with $m = 50$ gives a probability of success of 0.97. Comparatively, this same probability for the airport network is just 0.51. This difference is due to the ratio between $L$ and $M$ for each graph. As such, we can use a much larger sample size with the power-grid network.

Indeed, we could further increase the sample size here without reducing the probability of success too much, though there is not much to be gained in practical terms by doing this as the absolute time taken to successfully rewire this graph is quite low. That being said, one could calculate from Equation~(\ref{eqn:probsuccess}) an upper bound on the sample size that should be used for a given network, though this also depends on the distance between the starting assortativity value and the target assortativity, as discussed earlier. 

\subsection{Scalability to larger graphs}
\label{results:largegraphs}
As a final experiment, we test both algorithms on the Dogster network, which has approximately $250,000$ nodes and $2$ million edges. As we can see in Figure~\ref{fig:dogster_positive}, our algorithm reaches the target before there is a noticeable difference in the current assortativity value using the original algorithm, showing that our algorithm remains preferable in the case of large and dense networks. Note that for this graph, the minimum and maximum possible assortativity values were $-0.113$ and $0.048$ respectively, hence the target assortativity value is very close to the initial value.

To note, all of the experiments described in this section are repeated for the case of rewiring the network to be disassortative. For the sake of brevity, we only discuss the case of positive rewiring here, as the results in the negative rewiring case are almost identical, with the only differences being what choice of $m$ leads to the fastest rewiring for each network. Plots of these results corresponding to Figures~\ref{fig:airportresults},~\ref{fig:deezer_positive}, and~\ref{fig:dogster_negative} can be found in the~\hyperref[section:appendix]{Appendix}, as well as results for the randomly generated graphs (Figure~\ref{fig:RandomGraphs}).

We note also a possible limitation to this work. Given that we impose only that the degrees of nodes remains unchanged, it is likely that drastic changes in other network properties occur during the rewiring process, some of which may be undesirable. One can easily imagine that the clustering coefficient of a network is dramatically affected by rewiring. As we will see in the next section, holding other network properties fixed during the rewiring process is one of many potential avenues for further research.
\section{conclusions}
\label{section:conclusions}
In this paper we have introduced the FTL rewiring algorithm for tuning the assortativity of a network via edge rewiring. We demonstrate that this new version reduces computation time by several orders of magnitude, performing even better in the most complex case of rewiring large, densely connected graphs to assortativity values that are far from the starting value. 

We touched on the maximum and minimum assortativity values that can be realised by a graph with a given degree sequence in Section~\ref{section:results}, and we believe further exploration of this to be an interesting avenue for future work. Additionally we aim to formulate a way to obtain an optimum for the number of edges, $m$, for a graph based on its number of nodes, edges, and perhaps its degree distribution. Furthermore, we believe that the concepts used to develop this algorithm could be applied to other objectives, such as changing the level of clustering in a graph, as well as more intricate measurements of complex networks, such as fault-tolerance or robustness. This would also address the limitations to this study mentioned in Section~\ref{results:largegraphs}.

We have provided a Python package implementing this algorithm that can be used by anyone. This package can be found on GitHub at the link provided in the~\hyperref[section:appendix]{Appendix}.

As rewiring is an important part of many models used in network science across a wide range of areas of study, we expect that the approach presented in this paper will prove useful for the network science community.

\section*{Acknowledgements}
The authors would like to thank Piet Van Meighem and co-authors for access to the data used in \cite{van2010influence}, which was used to compare the performance of the different algorithms. We thank The networks research group at the University of Limerick as well as Laurent H\'ebert-Dufresne and Yingyue Ke for helpful conversations and discussions of the problem.
\section*{Funding}
This publication has emanated from research supported in part by a grant from Science Foundation Ireland under Grant number 18/CRT/6049. For the purpose of Open Access, the author has applied a CC BY public copyright licence to any Author Accepted Manuscript version arising from this submission.

\bigskip

\section*{Appendix}\label{section:appendix}
\subsection*{Code availability}
The package containing our degree-preserving rewiring algorithm can be found at: \\
\url{https://github.com/shmannion/degree\_preserving\_rewiring}
\subsection{Additional plots}
The following plots correspond to negative versions of the experiments described in the main text.

\begin{figure*}[t]
    \centering
    \includegraphics[page=17, width=0.3\textwidth]{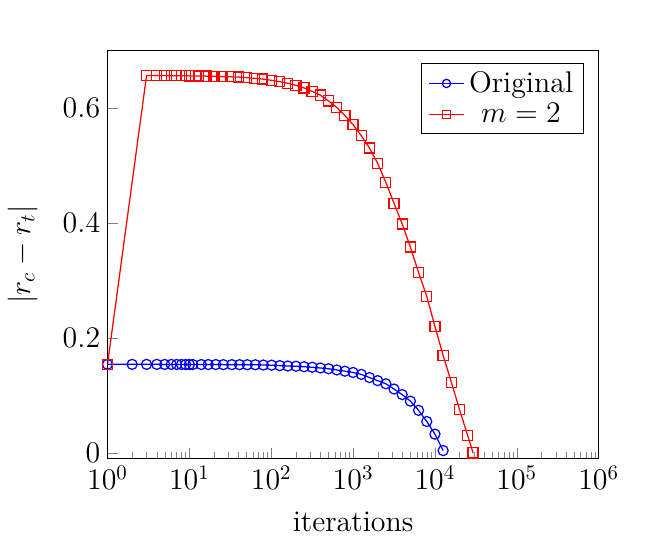}
    \includegraphics[page=18, width=0.3\textwidth]{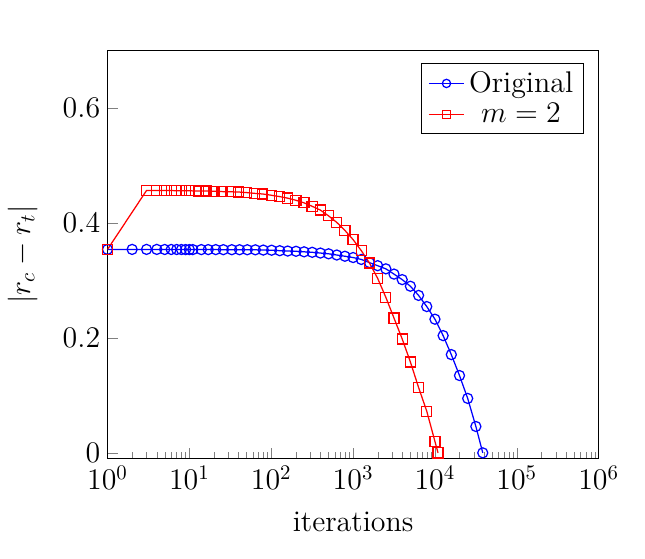}
    \includegraphics[page=19, width=0.3\textwidth]{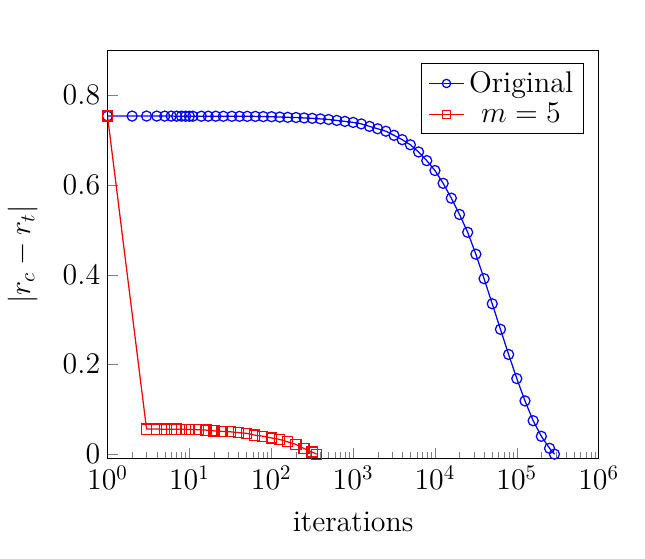}
    \includegraphics[page=20, width=0.3\textwidth]{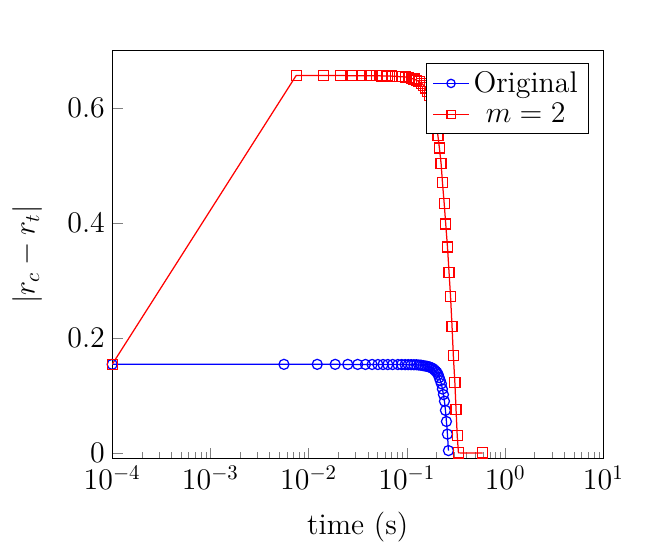}
    \includegraphics[page=21, width=0.3\textwidth]{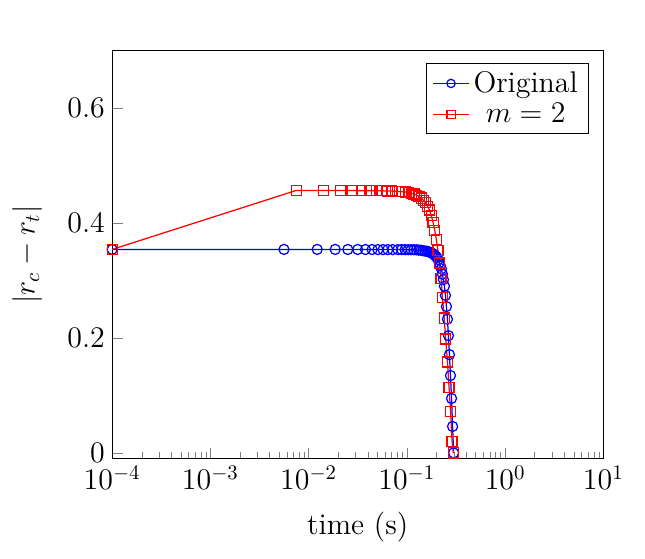}
    \includegraphics[page=22, width=0.3\textwidth]{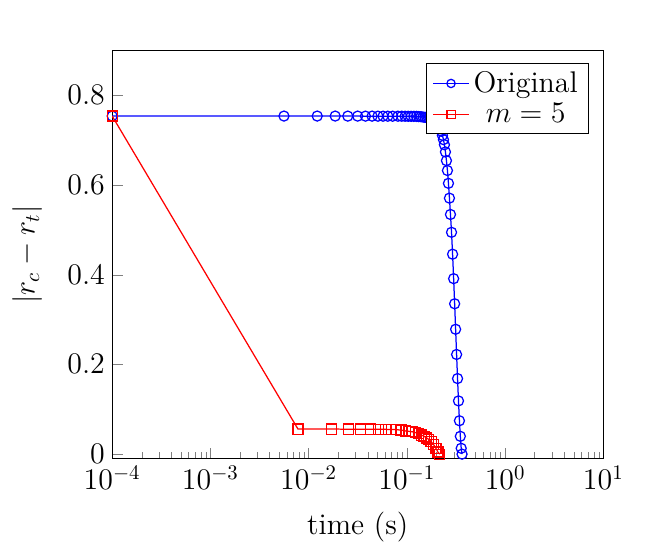}
    \caption{Difference between the current assortativity at a given iteration (top) or time (bottom) and the target assortativity for the US airport network. The target assortativity values are -0.2 (left), -0.4 (middle), and -0.8 (right). }
    \label{fig:airportresults_neg}
\end{figure*}

\begin{figure*}[h]
    \centering
    \includegraphics[page=75, width=0.3\textwidth]{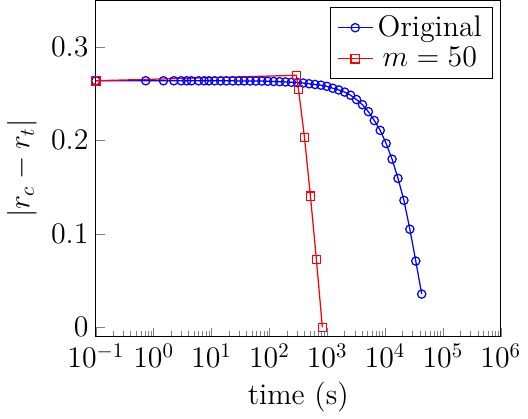}
    \includegraphics[page=87, width=0.3\textwidth]{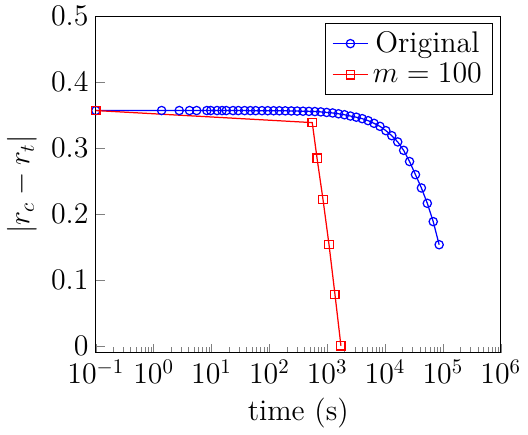}
    \includegraphics[page=79, width=0.3\textwidth]{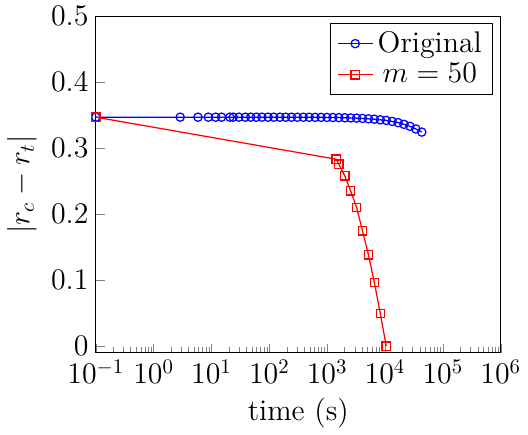}
    \includegraphics[page=76, width=0.3\textwidth]{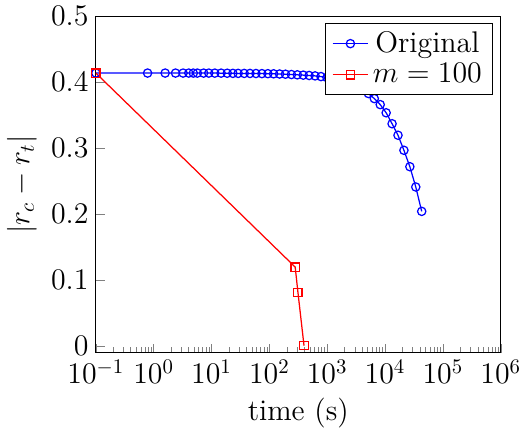}
    \includegraphics[page=88, width=0.3\textwidth]{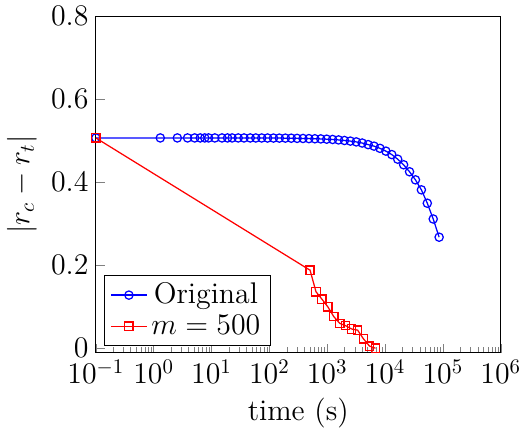}
    \includegraphics[page=80, width=0.3\textwidth]{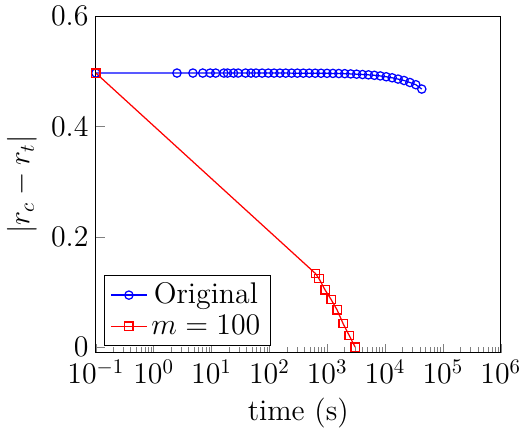}
    \caption{Difference between the current assortativity at a given time and the target assortativity for the different samples of the Deezer network; Romania (left), Hungary (middle), and Croatia (right). The target assortativity values are 0.2 (top) and 0.4 (bottom). Blue lines represent the original algorithm, and red points represent the FTL algorithm with the best performing choice of $m$. Note that we do not plot every time point to avoid visual clutter, hence the few markers for some curves. The original rewiring algorithm was stopped after 12 hours for the Romania and Croatia networks, and after 24 hours for the Hungary network. Hence, not all curves reach zero.}
    \label{fig:deezer_negative}
\end{figure*}

Figures~\ref{fig:airportresults_neg} and~\ref{fig:deezer_negative} show the results of using the original algorithm and the FTL algorithm to reduce the assortativity values of the networks. As we can see, the FTL algorithm once again outperforms the original.

\begin{figure*}[t]
    \centering
    \includegraphics[page=45, width=0.19\textwidth]{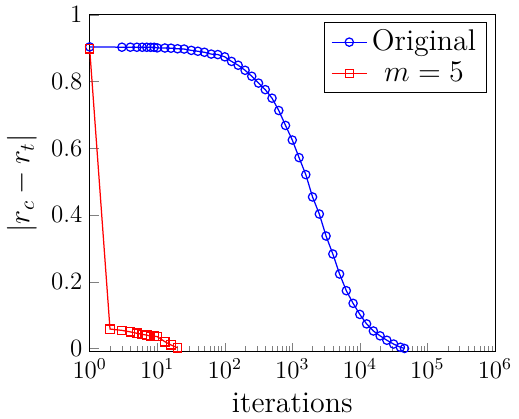}
    \includegraphics[page=47, width=0.19\textwidth]{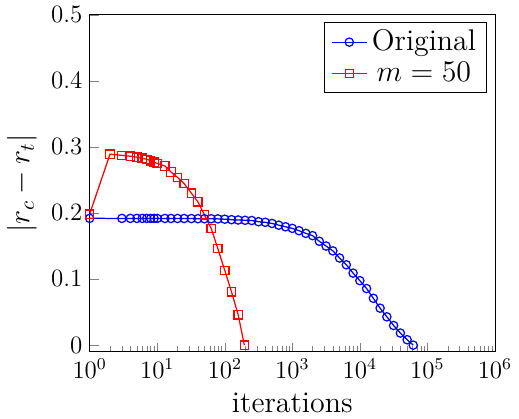}
    \includegraphics[page=43, width=0.19\textwidth]{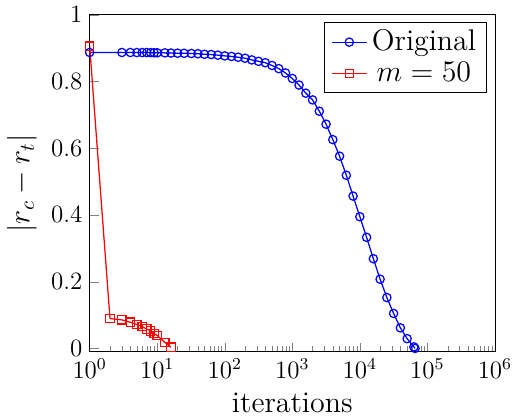}
    \includegraphics[page=49, width=0.19\textwidth]{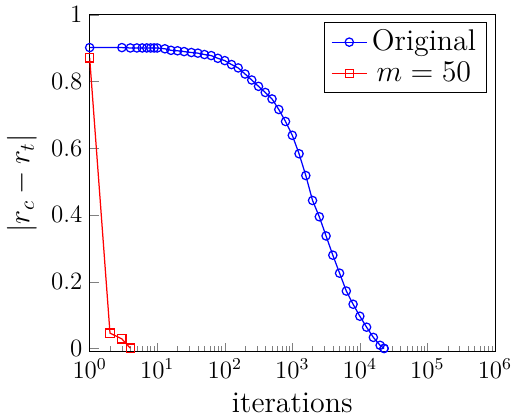}
    \includegraphics[page=51, width=0.19\textwidth]{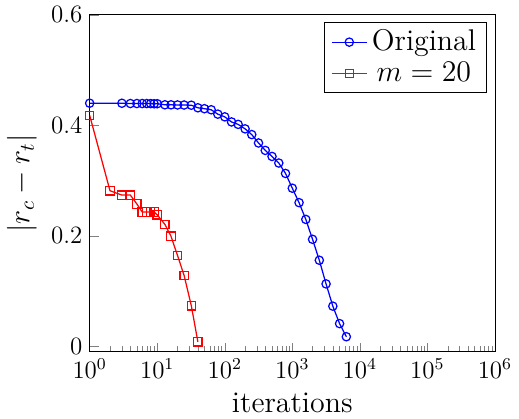}
    \includegraphics[page=46, width=0.19\textwidth]{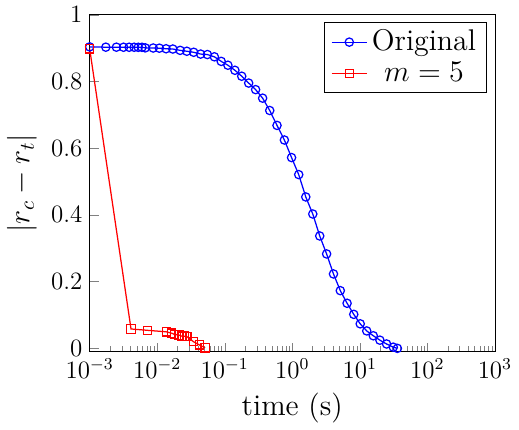}
    \includegraphics[page=48, width=0.19\textwidth]{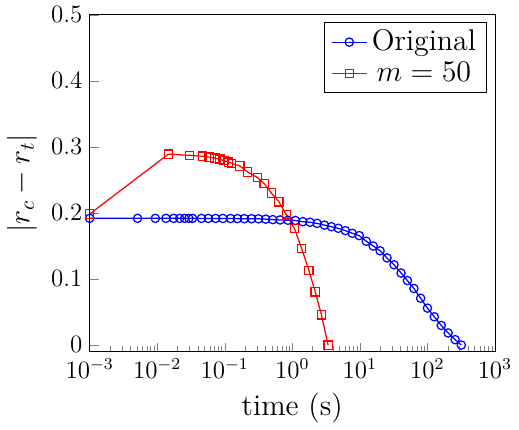}
    \includegraphics[page=44, width=0.19\textwidth]{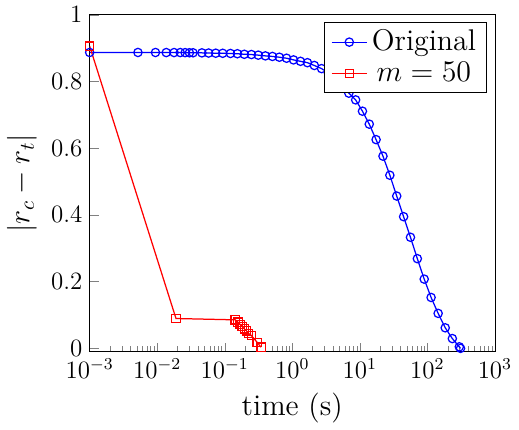}
    \includegraphics[page=50, width=0.19\textwidth]{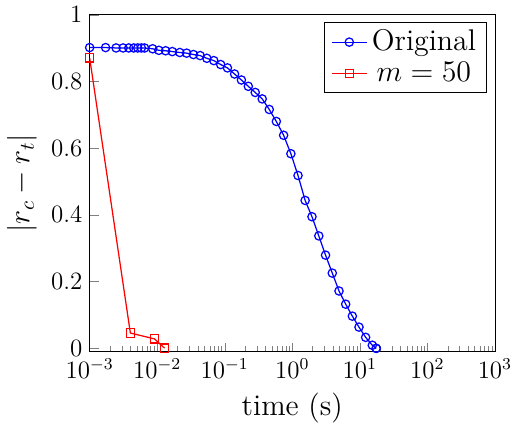}
    \includegraphics[page=52, width=0.19\textwidth]{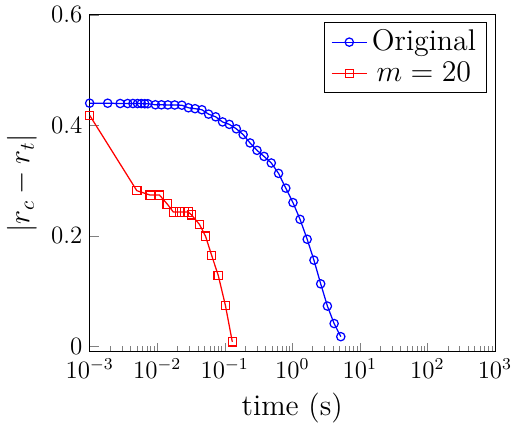}
    \caption{Performance of the original algorithm and the FTL algorithm on randomly generated graphs from the following distributions (from left to right): exponential ($r_t$ = 0.9), poisson (i.e., Erd\H{o}s-R\'enyi graphs, $r_t$ = 0.9), lognormal ($r_t$ = 0.85), power-law (i.e., Barab\'asi-Albert graphs, $r_t$ = 0.15), and Weibull ($r_t$ = 0.4). Blue lines represent the original algorithm, and red points represent the FTL algorithm with the best performing choice of $m$. }
    \label{fig:RandomGraphs}
    \vspace{1cm}
    \includegraphics[page=55, width=0.19\textwidth]{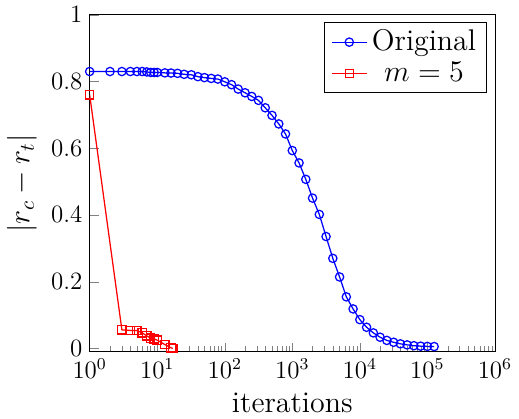}
    \includegraphics[page=59, width=0.19\textwidth]{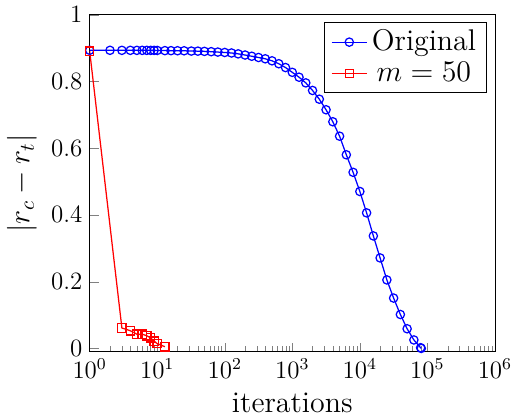}
    \includegraphics[page=57, width=0.19\textwidth]{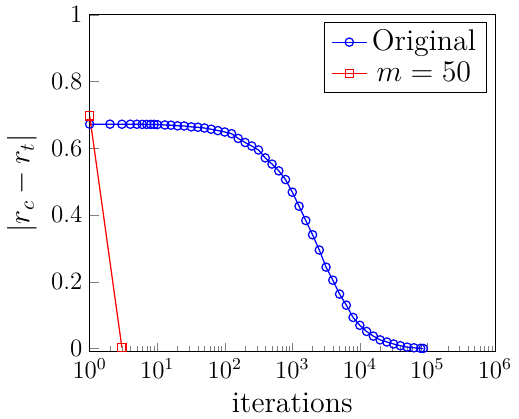}
    \includegraphics[page=61, width=0.19\textwidth]{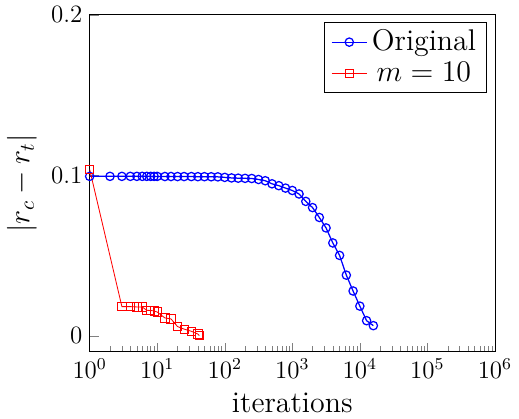}
    \includegraphics[page=63, width=0.19\textwidth]{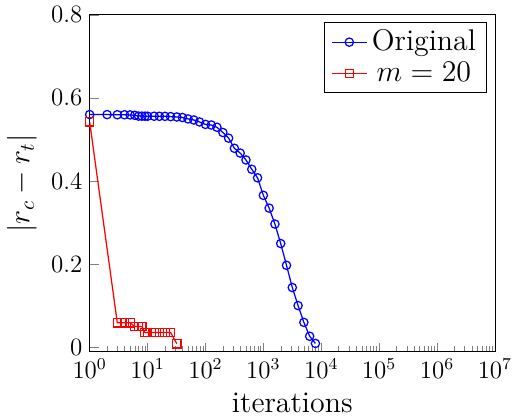}
    \includegraphics[page=56, width=0.19\textwidth]{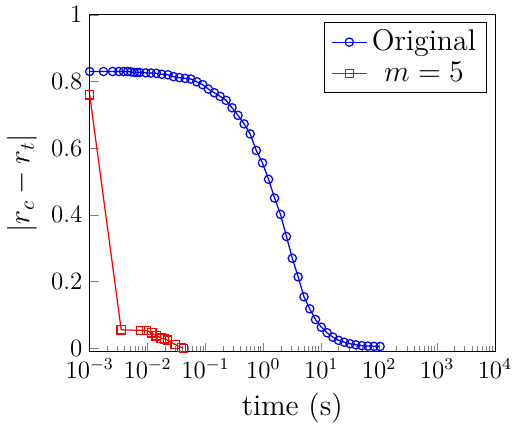}
    \includegraphics[page=60, width=0.19\textwidth]{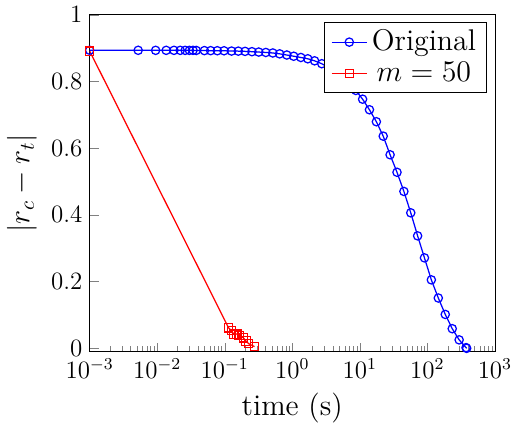}
    \includegraphics[page=58, width=0.19\textwidth]{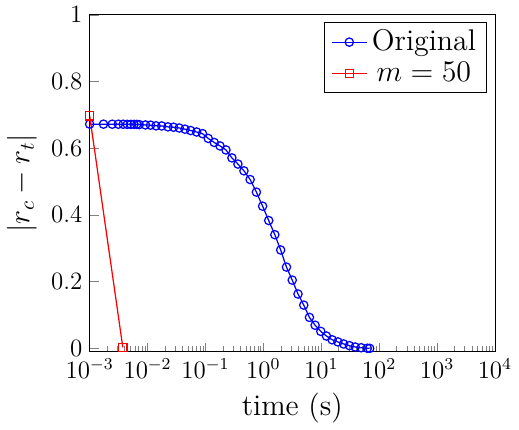}
    \includegraphics[page=62, width=0.19\textwidth]{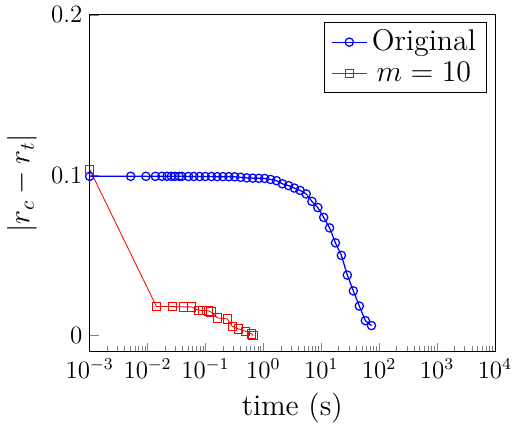}
    \includegraphics[page=64, width=0.19\textwidth]{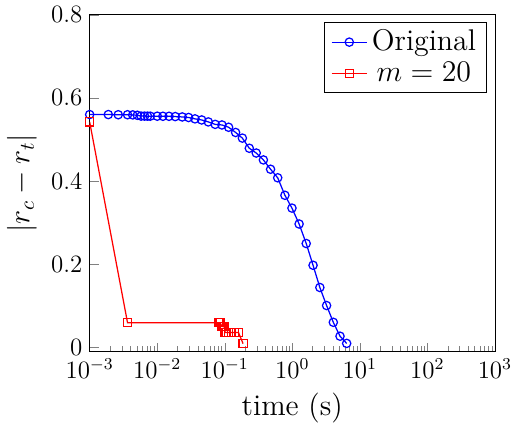}
    \caption{Performance of the original algorithm and the FTL algorithm on randomly generated graphs from the following distributions (from left to right): exponential ($r_t$ = -0.85), poisson (i.e., Erd\H{o}s-R\'enyi graphs, $r_t$ = -0.9), lognormal ($r_t$ = -0.7), Barab\'asi-Albert graphs ($r_t$ = -0.28), and Weibull ($r_t$ = -0.7). Blue lines represent the original algorithm, and red points represent the FTL algorithm with the best performing choice of $m$. }
    \label{fig:RandomGraphsNeg}
\end{figure*}

Finally, Figures~\ref{fig:RandomGraphs} and~\ref{fig:RandomGraphsNeg} show the performance of both algorithms on randomly generated graphs with different degree distributions. As we can see, regardless of the degree distribution of the graph, the FTL algorithm performs better.
\begin{figure*}
\centering

\begin{minipage}[t]{0.48\textwidth}
\hrule
\vspace{2pt}
\caption{Positive Degree-Preserving Rewiring}
\vspace{2pt}
\hrule
\begin{algorithmic}[1]
\Require Graph $G$, target $r$, edges per iteration $m$
\Return Rewired graph $G'$
\State $G' \gets G$
\State $nodes \gets$ nodes of $G$
\State $U \gets$ sort $nodes$ by descending original degree
\State $V \gets$ sort $nodes$ by descending original degree
\State remove all edges from $G'$

\For{node $u$ in $U$}
    \State $k \gets$ original degree of $u$
    \While{current degree of $u < k$}
        \For{node $v$ in $V$}
            \State $j \gets$ original degree of $v$
            \If{$u \neq v$}
                \If{degree($v$) $< j$}
                    \State add edge $(u,v)$ to $G'$
                \EndIf
            \EndIf
        \EndFor
    \EndWhile
    \State sort $V$ by descending remaining degree
\EndFor

\While{Assortativity($G'$) $> r$}
    \State $E \gets m$ random edges from $G'$
    \State $N \gets$ nodes in $E$
    \State sort $N$ by ascending degree
    \State $L \gets$ empty list of potential edges

    \For{$i$ in $(0,1,\dots,m)$}
        \State $u \gets N[i]$
        \State $v \gets N[2m - 1 - i]$
        \If{$u \neq v$}
            \State add edge $(u, v)$ to $L$
        \EndIf
    \EndFor

    \State check $L$ for repeat edges and existing edges
    \If{all edges in $L$ are valid}
        \State remove edges $E$ from $G'$
        \State add edges $L$ to $G'$
    \EndIf
\EndWhile

\State \Return $G'$
\end{algorithmic}
\hrule
\end{minipage}
\hfill
\begin{minipage}[t]{0.48\textwidth}
\hrule
\vspace{2pt}
\caption{Negative Degree-Preserving Rewiring}
\vspace{2pt}
\hrule
\begin{algorithmic}[1]

\Require Graph $G$, target $r$, edges per iteration $m$
\Return Rewired graph $G'$
\State $G' \gets G$
\State $nodes \gets$ nodes of $G$
\State $U \gets$ sort $nodes$ by ascending original degree
\State $V \gets$ sort $nodes$ by descending original degree

\State remove all edges from $G'$
\For{node $u$ in $U$}
    \State $k \gets$ original degree of $u$
    \While{current degree of $u < k$}
        \For{node $v$ in $V$}
            \State $j \gets$ original degree of $v$
            \If{$u \neq v$}
                \If{degree($v$) $< j$}
                    \State add edge $(u,v)$ to $G'$
                \EndIf
            \EndIf
        \EndFor
    \EndWhile
    \State sort $V$ by descending remaining degree
\EndFor

\While{Assortativity($G'$) $< r$}
    \State $E \gets m$ random edges from $G'$
    \State $N \gets$ nodes in $E$
    \State sort $N$ by ascending degree

    \For{$i$ in $(0,1,\dots,2m-2)$}
        \State $L \gets$ empty list of potential edges
        \State $u \gets N[i]$
        \State $v \gets N[i+1]$
        \If{$u \neq v$}
            \State add edge $(u, v)$ to $L$
        \EndIf
    \EndFor

    \State check $L$ for repeat edges and existing edges
    \If{all edges in $L$ are valid}
        \State remove edges $E$ from $G'$
        \State add edges $L$ to $G'$
    \EndIf
\EndWhile

\State \Return $G'$
\end{algorithmic}
\hrule
\end{minipage}

\end{figure*}
\bibliography{ref} 

@article{van2010influence,
  title={Influence of assortativity and degree-preserving rewiring on the spectra of networks},
  author={Van Mieghem, Piet and Wang, Huijuan and Ge, Xin and Tang, Siyu and Kuipers, Fernando A},
  journal={The European Physical Journal B},
  volume={76},
  number={4},
  pages={643--652},
  year={2010},
  publisher={Springer}
}

@article{maccarron2023correlation,
  title={Correlation distances in social networks},
  author={MacCarron, P{\'a}draig and Mannion, Shane and Platini, Thierry},
  journal={Journal of Complex Networks},
  volume={11},
  number={3},
  year={2023},
  publisher={Oxford University Press}
}

@article{chan2016rewiringforrobustness,
  title={Optimizing network robustness by edge rewiring: a general framework},
  author={Chan, Hau and Akoglu, Leman},
  journal={Data Mining and Knowledge Discovery},
  volume={30},
  pages={1395--1425},
  year={2016},
  publisher={Springer}
}

@article{winterbach2012greedy,
  title={Do greedy assortativity optimization algorithms produce good results?},
  author={Winterbach, Wynand and de Ridder, Dick and Wang, HJ and Reinders, Marcel and Van Mieghem, Piet},
  journal={The European Physical Journal B},
  volume={85},
  pages={1--9},
  year={2012},
  publisher={Springer}
}

@article{kashyap2018mechanisms,
  title={Mechanisms for tuning clustering and degree-correlations in directed networks},
  author={Kashyap, G and Ambika, G},
  journal={Journal of Complex Networks},
  volume={6},
  number={5},
  pages={767--787},
  year={2018},
  publisher={Oxford University Press}
}

@article{bertotti2020rkplane,
  title={Network Rewiring in the r-K Plane},
  author={Bertotti, Maria Letizia and Modanese, Giovanni},
  journal={Entropy},
  volume={22},
  number={6},
  pages={653},
  year={2020},
  publisher={MDPI}
}

@article{alstott2019localclustering,
  title={Local rewiring algorithms to increase clustering and grow a small world},
  author={Alstott, Jeff and Klymko, Christine and Pyzza, Pamela B and Radcliffe, Mary},
  journal={Journal of Complex Networks},
  volume={7},
  number={4},
  pages={564--584},
  year={2019},
  publisher={Oxford University Press}
}

@article{bertsimas1993simulatedannealing,
  title={Simulated annealing},
  author={Bertsimas, Dimitris and Tsitsiklis, John},
  journal={Statistical science},
  volume={8},
  number={1},
  pages={10--15},
  year={1993},
  publisher={Institute of Mathematical Statistics}
}

@inproceedings{buesser2011optimizingannealing,
  title={Optimizing the robustness of scale-free networks with simulated annealing},
  author={Buesser, Pierre and Daolio, Fabio and Tomassini, Marco},
  booktitle={Adaptive and Natural Computing Algorithms: 10th International Conference, ICANNGA 2011, Ljubljana, Slovenia, April 14-16, 2011, Proceedings, Part II 10},
  pages={167--176},
  year={2011},
  organization={Springer}
}

@article{de2018fundamentals,
  title={Fundamentals of spreading processes in single and multilayer complex networks},
  author={de Arruda, Guilherme Ferraz and Rodrigues, Francisco A and Moreno, Yamir},
  journal={Physics Reports},
  volume={756},
  pages={1--59},
  year={2018},
  publisher={Elsevier}
}

@techreport{networkx,
  title={Exploring network structure, dynamics, and function using NetworkX},
  author={Hagberg, Aric and Swart, Pieter and S Chult, Daniel},
  year={2008},
  institution={Los Alamos National Lab.(LANL), Los Alamos, NM (United States)}
}

@book{newman2018networks,
  title={Networks},
  author={Newman, Mark},
  year={2018},
  publisher={Oxford university press}
}

@article{mannion2023robust,
  title={A robust method for fitting degree distributions of complex networks},
  author={Mannion, Shane and MacCarron, P{\'a}draig},
  journal={Journal of Complex Networks},
  volume={11},
  number={4},
  pages={},
  year={2023},
  publisher={Oxford University Press}
}

@article{newman2002assortative,
  title={Assortative mixing in networks},
  author={Newman, Mark EJ},
  journal={Physical review letters},
  volume={89},
  number={20},
  pages={208701},
  year={2002},
  publisher={APS}
}

@article{albert2002statistical,
  title={Statistical mechanics of complex networks},
  author={Albert, R{\'e}ka and Barab{\'a}si, Albert-L{\'a}szl{\'o}},
  journal={Reviews of modern physics},
  volume={74},
  number={1},
  pages={47},
  year={2002},
  publisher={APS}
}

@article{amaral2000classes,
  title={Classes of small-world networks},
  author={Amaral, Lu{\i}s A Nunes and Scala, Antonio and Barthelemy, Marc and Stanley, H Eugene},
  journal={Proceedings of the national academy of sciences},
  volume={97},
  number={21},
  pages={11149--11152},
  year={2000},
  publisher={National Acad Sciences}
}

@article{newman2001random,
  title={Random graphs with arbitrary degree distributions and their applications},
  author={Newman, Mark EJ and Strogatz, Steven H and Watts, Duncan J},
  journal={Physical review E},
  volume={64},
  number={2},
  pages={026118},
  year={2001},
  publisher={APS}
}

@article{hakimi1962realizability,
  title={On realizability of a set of integers as degrees of the vertices of a linear graph. I},
  author={Hakimi, S Louis},
  journal={Journal of the Society for Industrial and Applied Mathematics},
  volume={10},
  number={3},
  pages={496--506},
  year={1962},
  publisher={SIAM}
}

@article{havel1955remark,
  title={A remark on the existence of finite graphs},
  author={Havel, V{\'a}clav},
  journal={Casopis Pest. Mat.},
  volume={80},
  pages={477--480},
  year={1955}
}

@article{maslov2002originalrewiring,
  title={Specificity and stability in topology of protein networks},
  author={Maslov, Sergei and Sneppen, Kim},
  journal={Science},
  volume={296},
  number={5569},
  pages={910--913},
  year={2002},
  publisher={American Association for the Advancement of Science}
}

@article{SHANG201749,
title = {Fitness networks for real world systems via modified preferential attachment},
journal = {Physica A: Statistical Mechanics and its Applications},
volume = {474},
pages = {49-60},
year = {2017},
author = {Ke-ke Shang and Michael Small and Wei-sheng Yan}
}

@article{xulvi2004reshuffling,
  title={Reshuffling scale-free networks: From random to assortative},
  author={Xulvi-Brunet, Ramon and Sokolov, Igor M},
  journal={Physical Review E—Statistical, Nonlinear, and Soft Matter Physics},
  volume={70},
  number={6},
  pages={066102},
  year={2004},
  publisher={APS}
}

@article{stokes2018common,
  title={Common greedy wiring and rewiring heuristics do not guarantee maximum assortative graphs of given degree},
  author={Stokes, Jonathan and Weber, Steven},
  journal={Information processing letters},
  volume={139},
  pages={53--59},
  year={2018},
  publisher={Elsevier}
}

@article{meghanathan2016maximal,
  title={Maximal assortative matching for complex network graphs},
  author={Meghanathan, Natarajan},
  journal={Journal of King Saud University-Computer and Information Sciences},
  volume={28},
  number={2},
  pages={230--246},
  year={2016},
  publisher={Elsevier}
}

@inproceedings{konect:all,
  title={Konect: the koblenz network collection},
  author={Kunegis, J{\'e}r{\^o}me},
  booktitle={Proceedings of the 22nd International Conference on World Wide Web},
  pages={1343--1350},
  year={2013}
}

@inproceedings{deezer,    
    title={GEMSEC: Graph Embedding with Self Clustering},    
    author={Rozemberczki, Benedek and Davies, Ryan and Sarkar, Rik and Sutton, Charles},    
    booktitle={Proceedings of the 2019 IEEE/ACM International Conference on Advances in Social Networks Analysis and Mining 2019},    
    pages={65-72},    
    year={2019},    
    organization={ACM}    
}
\end{document}